\documentclass[conference]{IEEEtran}
\IEEEoverridecommandlockouts

\usepackage{cite}
\usepackage{amsmath,amssymb,amsfonts}
\usepackage{algorithmic}
\usepackage{graphicx}
\usepackage{textcomp}
\usepackage{xcolor}
\usepackage{tcolorbox}

\usepackage[T1]{fontenc}
\usepackage[utf8]{inputenc}
\usepackage{tabularx,ragged2e,booktabs,caption}
\newcolumntype{C}[1]{>{\Centering}m{#1}}

\usepackage{caption}
\usepackage[utf8]{inputenc}
\usepackage[T1]{fontenc}
\usepackage{hyperref}
\usepackage{graphicx}
\usepackage{multirow}
\usepackage{graphics} 
\usepackage{epsfig} 
\usepackage{mathptmx} 
\usepackage{mathptmx} 
\usepackage{amsmath} 
\usepackage{amssymb}  

\def\BibTeX{{\rm B\kern-.05em{\sc i\kern-.025em b}\kern-.08em
    T\kern-.1667em\lower.7ex\hbox{E}\kern-.125emX}}
\begin{document}

\makeatletter
\newcommand{\linebreakand}{%
  \end{@IEEEauthorhalign}
  \hfill\mbox{}\par
  \mbox{}\hfill\begin{@IEEEauthorhalign}
}

\makeatother
\title{\bf EEG aided boosting of single-lead ECG based sleep staging with Deep Knowledge Distillation\\
}
\author{
\IEEEauthorblockN{Vaibhav Joshi}
\IEEEauthorblockA{\textit{Department of Electrical Engineering} \\
\textit{IIT Madras}\\
Chennai, India \\
ee19s039@smail.iitm.ac.in}
\and
\IEEEauthorblockN{Sricharan V}
\IEEEauthorblockA{\textit{Healthcare Technology Innovation} \\
\textit{Innovation Center, IIT Madras}\\
Chennai, India \\
sricharanv@htic.iitm.ac.in}
\linebreakand
\IEEEauthorblockN{Preejith SP}
\IEEEauthorblockA{\textit{Healthcare Technology Innovation} \\
\textit{Innovation Center, IIT Madras}\\
Chennai, India \\
preejith@htic.iitm.ac.in}
\and
\IEEEauthorblockN{Mohanasankar Sivaprakasam}
\IEEEauthorblockA{\textit{Department of Electrical Engineering} \\
\textit{IIT Madras}\\
Chennai, India \\
mohan@ee.iitm.ac.in}
}
\maketitle

\begin{abstract}
An electroencephalogram (EEG) signal is currently accepted as a standard for automatic sleep staging. Lately, Near-human accuracy in automated sleep staging has been achievable by Deep Learning (DL) based approaches, enabling multi-fold progress in this area. However, An extensive and expensive clinical setup is required for EEG based sleep staging. Additionally, the EEG setup being obtrusive in nature and requiring an expert for setup adds to the inconvenience of the subject under study, making it adverse in the point of care setting. An unobtrusive and more suitable alternative to EEG is Electrocardiogram (ECG). Unsurprisingly, compared to EEG in sleep staging, its performance remains sub-par. In order to take advantage of both the modalities, transferring knowledge from EEG to ECG is a reasonable approach, ultimately boosting the performance of ECG based sleep staging. Knowledge Distillation (KD) is a promising notion in DL that shares knowledge from a superior performing but usually more complex teacher model to an inferior but compact student model. Building upon this concept, a cross-modality KD framework assisting features learned through models trained on EEG to improve ECG-based sleep staging performance is proposed.
Additionally, to better understand the distillation approach, extensive experimentation on the independent modules of the proposed model was conducted. Montreal Archive of Sleep Studies (MASS) dataset consisting of 200 subjects was utilized for this study. The results from the proposed model for weighted-F1-score in 3-class and 4-class sleep staging showed a 13.40 \% and 14.30 \% improvement, respectively. This study demonstrates the feasibility of KD for single-channel ECG based sleep staging's performance enhancement in 3-class (W-R-N) and 4-class (W-R-L-D) classification.

\end{abstract}
\section{Introduction}

\begin{figure}[t]
    \centering
      {
      \centering
      \includegraphics[width=0.45\textwidth]{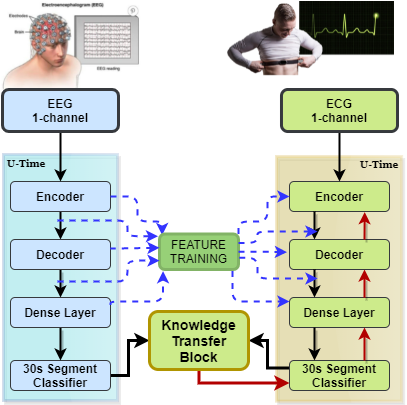}
}
    
    \caption{Proposed Knowledge Distillation framework}
    \label{fig:1}
\end{figure}

Sleep is an intricate dynamic physiological process that occurs in multi-cyclical stages. In sleep medicine, sleep is typically studied by acquiring multiple bio-signals during sleep by conducting a polysomnography (PSG) study. The primary reference for sleep studies is accepted to be the Electroencephalogram (EEG) signal, considering its interpretability with brain activation, the pivot of sleep mechanism. Generally, experts manually perform sleep stage classification for 30/20 s epochs following the sleep staging guidelines and rules by Rechtschaffen and Kales (1968) (the ‘R and K rules’) \cite{rechtschaffen1968manual} or AASM (American Academy of Sleep Medicine) \cite{iber2007aasm}. As manual sleep staging is cumbersome and time-consuming, many automated sleep staging algorithms, including neural network-based approaches \cite{perslev2019u}, have been developed lately with notable performance on par with human accuracy. Traditionally, sleep is divided into five stages:  W, wakefulness; REM Rapid Eye Movement stage; N1, a light sleep period in Non-REM stages; N2, an intermediate stage; N3, a deep sleep stage. Different frequencies and patterns observed in the EEG signal during sleep characterize different sleep stages. The Color Density Spectral Array (CDSA), as shown in Fig.\ref{fig:2} shows EEG and ECG frequency components for different sleep stages. The patterns match in both signals, with the EEG pattern being more distinctive for sleep stages. However,  in a non-clinical setup, the obtrusiveness of EEG renders it impractical. Furthermore, during sleep, the brain-body interaction implies that the stages of sleep can be captured by other physiological signals as well \cite{abdullah2009correlation}.

\begin{figure}[t]
    \centering
      {
      \centering
      \includegraphics[width=0.45\textwidth]{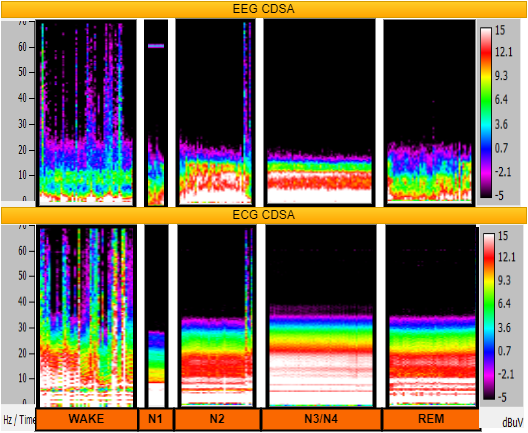}
}
    
    \caption{Color Density Spectral Array-CDSA of EEG and ECG}
    \label{fig:2}
\vspace{-1.2em}
\end{figure}

An electrocardiogram (ECG), not only being readily integrable into wearable devices but also less obtrusive, is an appropriate alternative to the EEG. As seen in Fig.\ref{fig:2}, the ECG pattern is not distinctive, So Deep Learning approach is a plausible choice. Q.Li \textit{et al.} \cite{li2018deep} extracted Respiratory Sinus Arrhythmia (RSA) and ECG Derived Respiration (EDR) based features from a vast PSG dataset (SHHS, CinC, SLPDB) for cross-spectral spectrogram, which helped to generalize the model better and achieved significant results for sleep staging using ECG signal. Their work employed an SVM, which was given high-level features extracted from the input using a Convolutional Neural Network(CNN). An accuracy of 65.90 \% on SHHS and 75.40 \% on SLPDB, respectively, were obtained for 4-class sleep staging. Radha \textit{et al.} \cite{radha2019sleep} used a temporal model approach, LSTM for sleep staging on 132 HRV features explicitly extracted from ECG signal and achieved an accuracy of 77.0 \%. In another study, Fonseca \textit{et al.} \cite{fonseca2015sleep} extracted 80 expert features from Respiratory Inductance Plethysmography (RIP)and ECG eventually used by Linear Discriminant (LD), which obtained an accuracy of 80.0 \% and 69.0 \% for 3-class and 4-class sleep staging, respectively. In a separate study by Sridhar \textit{et al.}, \cite{sridhar2020deep}, 2 stage CNNs achieved an accuracy of 77.0 \% from ECG-derived heart rate on an extensive dataset comprising MESA, CinC and SHHS. Despite the burgeoning potential of ECG, EEG based sleep classification performance remains vastly superior for sleep staging \cite{perslev2019u}\cite{sun2020sleep}. It would be exceedingly beneficial to integrate the advantages of improved accuracy with unobtrusive monitoring. While multi-modal fusion methods utilize features of multiple signals and have obtained improved accuracy \cite{fonseca2015sleep}\cite{sun2020sleep}, multiple signal acquisition is required during inference which is an additional overhead. Thus, using the primary modality, EEG, to impart information to ECG opens up a lot of potential applications.

 Knowledge Distillation (KD) has recently gained traction in deep networks to effectively transfer relevant information to more compact networks from extensive networks. A response based KD method to transfer and distil information to a student model from the softmax layer of a larger size teacher model to achieve better generalization was proposed by Hinton \textit{et al.} \cite{hinton2015distilling}. Another approach that implemented feature-based distillation through knowledge distillation from the intermediate feature maps instead of the softmax layers was proposed by Romero \textit{et al.} \cite{romero2014fitnets}. The concept to transfer attention maps from the intermediate layers to improvise the feature distillation process was introduced by Zagoruyko \textit{et al.} \cite{komodakis2017paying}. Cross-modal KD \cite{gou2021knowledge} explored the above idea by applying it across modalities.
Inspired by these approaches, a cross-modal KD approach, as depicted in Fig.\ref{fig:1} is proposed. The proposed method combines feature-based and response based distillation facilitating multi-modal training of ECG and EEG while enabling uni-modal testing. This study is designed to evaluate the potency of KD in the performance enhancement of ECG based sleep staging. Moreover, we compare individual elements of the KD framework and the proposed model, subsequently exhibiting its potential.
 

\section{METHODS}

\subsection{Problem Formulation}

Let $s_{eeg}$ $ \in\ \mathbb{R}^{t f} $\ and $s_{ecg}$ $ \in\ \mathbb{R}^{ t f} $\ be the EEG and ECG waveforms, with a sampling rate \textit{f} for $ t $ seconds, respectively. Let $M(eeg; \phi_{eeg})$ and $M(ecg; \phi_{ecg})$ be the models which take T
connected segments, each having length \textit{i}, from $s_{eeg}$ and $s_{ecg}$ respectively. Let \textit{e} be the frequency at which the signal is split, where the aim is to independently match $s_{eeg}$ and $s_{ecg}$ to [\textit{e} * $t$] labels,
where each label is based on \textit{i= f/\textit{e}} sampled points. The model we adapt is able to handle varying frequencies during inference. To be more specific, the model $M(eeg; \phi_{eeg})$ and $M(ecg; \phi_{ecg})$
maps $s_{eeg}$ and $s_{ecg}$ to ground truth for predicting C
classes in all T segments.  

Weighted Cross Entropy(WCE) is the chosen loss function that is to be optimized by the teacher model, $M(eeg; \phi_{eeg})$ and is  defined as:

\begin{equation} \label{eq:1}
L(s_{eeg}, y) = \sum_{i=1}^{T} 1/\sum_{i=1}^{T} (-w_{yi})  * l_{eeg}^i
     \end{equation}

where $w_{yi}$ is the class weight which is proportional to the number of data points in the training set from a particular class  and

\begin{equation} \label{eq:2}
l_{eeg}^i = (log(exp(s_{eeg}^{(i,y_i)}) * w_{y_i}) / \sum_{k=1}^{C}  exp(s_{eeg}^{(i,k)})  
\end{equation}

This is followed by Feature Based (FB) distillation of the aggregated attention maps obtained from the features of the pretrained teacher model, $M(eeg; \phi_{eeg})$, to the untrained student model, $M(ecg; \phi_{ecg})$. To obtain optimal FB distillation, we adopt the following Attention Transfer (AT) loss \cite{komodakis2017paying}:

\begin{equation} \label{eq:3}
    L_{FB} = \sum_{j \in I}||\frac{Q_{ecg}^{j}}{||Q_{ecg}^{j}||_{2}} - \frac{Q_{eeg}^{j}}{||Q_{eeg}^{j}||_{2}}||_{2}
\end{equation}

where the $j$-th pair of teacher and student attention maps  is represented by $Q_{eeg}^{j}$=$vec(F_{eeg}(A_{eeg}^{j}))$ and $Q_{ecg}^{j}$=$vec(F_{ecg}(A_{ecg}^{j}))$ in a vectorized form, $I$ denote the set of teacher-student convolution layers which is selected for Attention Transfer. In our framework, $j$ distils the attention maps from all the layers by iterating through the features of the whole architecture.
In AT, the main motive is to train a student network that, while being accurate, will also have attention maps that are similar to those of the teacher.

\begin{figure*}[t]
    \centering
    {
      \centering
      \includegraphics[width=1\textwidth, height = 0.55\textwidth]{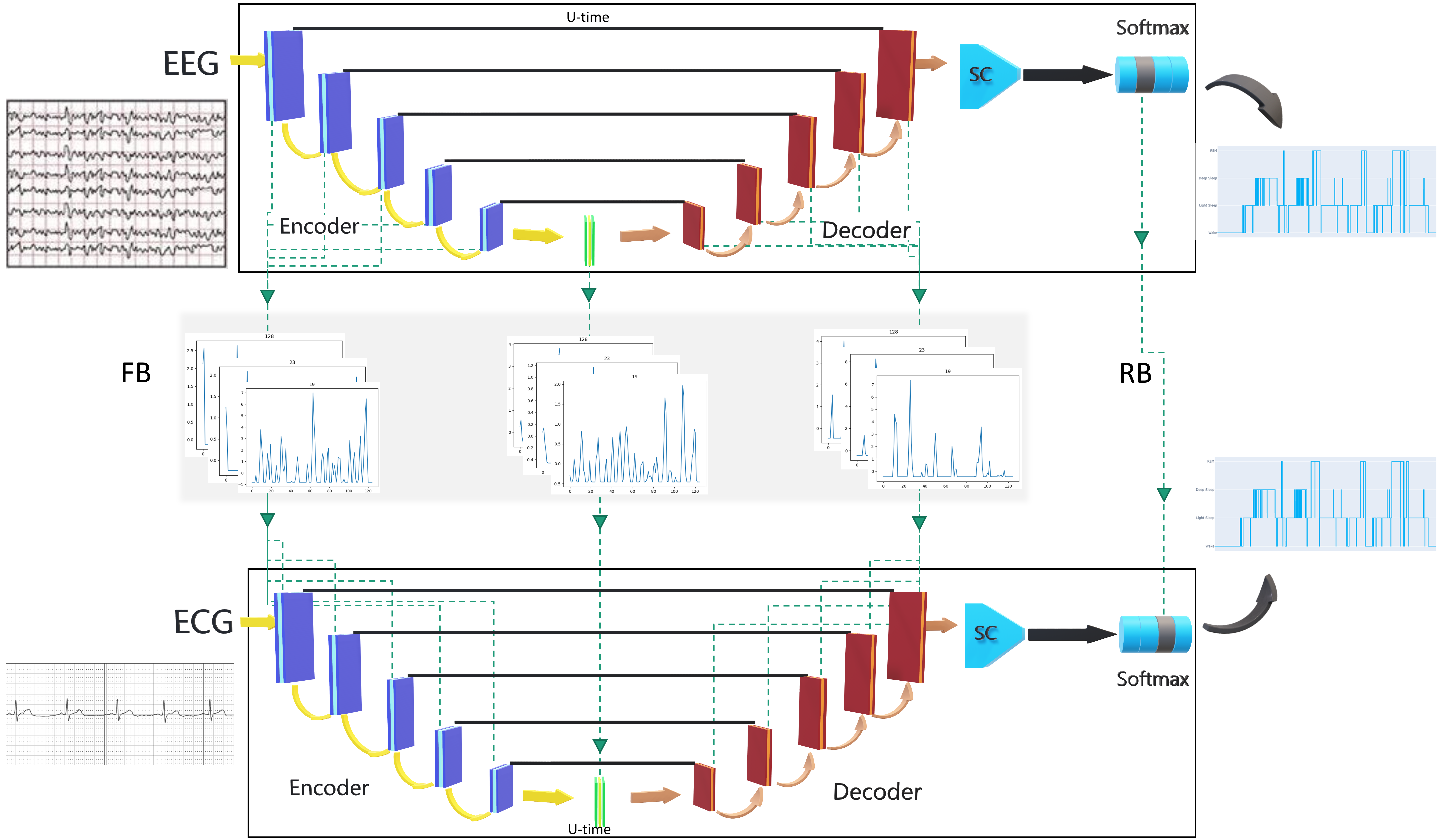}

}
    
    \caption{Knowledge transfer in action from EEG sleep staging model to ECG sleep staging model. SC: Segment Classifier from U-Time. FB: Feature based (Attention Transfer distillation) feature learning. RB: Response based (Softmax distillation) feature learning.}
    \label{fig:Feature_transfer}
\end{figure*}

Following this, the pretrained student model $M(ecg; \phi_{ecgpre})$, optimizes the sum of the Response Based (RB) distillation and WCE loss defined as:

\begin{equation} \label{eq:4}
L(s_{ecg}, y) = (1-\beta)  * l_{ecg}^i * \sum_{i=1}^{T} 1/\sum_{i=1}^{T} (-w_{yi}) +  \beta T_d^2 * l_{dist}^i
\end{equation}

where beta is the weight between RB and FB losses, $T_{s}$ is the temperature parameter of softmax and   
\begin{equation} \label{eq:5}
l_{ecg}^i = w_{y_i} * log(exp(s_{ecg}^{(i,y_i)}) / \sum_{k=1}^{C}  exp(s_{ecg}^{(i,k)})
\end{equation}

\begin{equation} \label{eq:6}
l_{dist}^i = KLD(p(s_{ecg}, T_s), p(s_{eeg}, T_s))
\vspace{-1.35em}
\end{equation}

\begin{equation} \label{eq:7}
p(x,T_s) = log(exp(x^{(i,y_i)}/T_s) / \sum_{k=1}^{C}  exp(x^{(i,k)}/T_s)
\end{equation}

Where $KLD$ denotes Kulback Leibler Divergence.

\subsection{Architectural Details}

The proposed model adopts the three module structure as proposed in the U-Time architecture \cite{perslev2019u} which includes an encoder, decoder and segment classifier. The encoder compresses the raw EEG or ECG signal into a group of subsampled feature maps, with the last layer acting as the bottleneck. It consists of five blocks where each block consists of two convolution subblocks followed by a max-pooling layer that subsamples the input by a factor of two. 
Finally, the bottleneck layer consists of two convolutions that retain the spatial size. The decoder is entasked to learn a mapping from the bottleneck layer back to the input signal domain giving out a dense segmentation map of the same size as the input. The resulting feature maps are concatenated with the corresponding feature maps computed by the encoder at the same scale. The five blocks in the decoder consist of two convolution subblocks, similar to the blocks in the encoder, followed by an upsampling layer.
The segment classifier is fed the output segmentation map to predict the final sleep stages at the desired resolution.

The above architecture serves as the base for both the teacher, $M(eeg; \phi_{eeg})$ and the student, $M(ecg; \phi_{ecg})$. This architecture was chosen considering following major reasons:
\begin{enumerate}
    \item \textbf{Fully convolutional architecture} - U-Time can be applied across any dataset without much architecture or hyperparameter tuning as it is fully convolutional.
    \item  \textbf{Customized for EEG} - U-Time was optimized to improve sleep staging from EEG. Picking this architecture will ensure that the best teacher model is chosen for feature transfer from EEG to ECG.
    \item  \textbf{Inference time variable-length segmentation} - A U-Time model may be used to stage sleep at any frequency, i.e., every 20 s or 40 s, at inference time.
    
\end{enumerate}


\subsection{Dataset Description}

This study utilizes the Montreal Archive of Sleep Studies (MASS) \cite{o2014montreal} dataset containing sleep recordings obtained from 200 participants [103 females (aged 38.3 ± 18.9 years) and 97 males (aged 42.9 ± 19.8 years); age range: 18–76 years]; organized into five sets of PSG records, SS1-SS5. The dataset was acquired online from the Centre for Advanced Research in Sleep Medicine (CARSM) on providing the project proposal approved by the local ethics board. All participants were part of a healthy control group, except for 15 of the SS1 subset who suffered from Mild Cognitive Impairment (MCI). The study utilizes data from all the 200 subjects. Among the many EEG electrodes positioned as per the international 10-20 system, the C3-A2/C4-A1 electrodes were used. Lead 1 was the electrode of choice from the ECG. The data was undersampled to 200 Hz from the initial sampling rate to optimize runtime and uniformity while comfortably satisfying the Nyquist criterion. A window width of thirty seconds was uniformly considered for training and inference. Subsets with annotation for every twenty seconds were converted into thirty-second segments by including five seconds of data before and after the annotation. The sleep stages N1 and N2  were combined into Light Sleep(L) and N3, and N4 into Deep Sleep(D) for the four-class (W-L-D-R)  classification problem. N1, N2, N3, and N4 were combined into a single NREM(N) stage for the three-class (W-N-R) classification problem.
 
\subsection{Experimental Procedures}

All the data was split subject-wise into train, eval and test sets in 80:10:10 ratio for both the classification problems. This ensured zero overlaps of data between splits from the same subject. The best model in all the runs was identified based on metrics tracked on the validation set during training. This was then validated on the holdout test-set post-training. The metrics used to validate the model's performance included accuracy and weighted F1 score \cite{perslev2019u}. The weighted-F1 score is calculated by computing the f1 score for each class separately and averaging the scores, weighted by individual class support (True positive + False Negative) to tackle the intrinsic imbalance in sleep staging. Experiments for both three and four class classification tasks were conducted identically as per the framework shown in Fig.\ref{fig:Feature_transfer}. Baselines were trained through the optimization of WCE loss as given in $Eq.\ref{eq:1}$ whereas distillation was carried out through the optimization of the loss given in $Eq.\ref{eq:4}$. Furthermore, we conducted two experiments to investigate the individual modules in our proposed framework. The primary component in all the experiments involve two significant steps as given below:
\begin{enumerate}
    \item Feature Training (Step 1): The loss ($Eq.\ref{eq:3}$) between ECG and EEG features is optimized by $M(ecg; \phi_{ecg})$ after freezing the EEG weights. This trains the ECG model to train towards imitating the feature maps of the EEG model.
    \item Final Training (Step 2): Subsequently, $M(ecg;\phi_{ecgpre})$ is trained further on RB loss($Eq.\ref{eq:4}$) for optimizing the ECG model weights. The T parameter was chosen as 1, as given in \cite{hinton2015distilling} and beta was chosen empirically to provide standard weights to classification and distillation loss.
\end{enumerate}

The implementation procedure of the proposed distillation method and ablation methods is as follows:

\begin{enumerate}
    \item \textbf{FB+RB+WCE(proposed method)}: Step 1 is executed to train features followed by training in step 2 on the loss $Eq.\ref{eq:4}$ with $ \beta = 0.5$.
    \item \textbf{FB+WCE(ablation method)}: Step 1 is executed to train features followed by training in step 2 on the loss $Eq.\ref{eq:4}$ with $ \beta = 0$ , which consequently trains independently on the classification loss  in $Eq.\ref{eq:1}$.
    \item \textbf{RB+WCE(ablation method)}: Only step 2 is executed training on the loss $Eq.\ref{eq:4}$ with $ \beta = 0.5$.
\end{enumerate}
\footnote{FB: Feature based; RB: Response based; WCE: Weighted Cross Entropy classification Loss}
Each configuration of distillation was trained with learning  rate(LR) of $10^{-3}$ for 150 epochs. Nvidia GTX3090Ti 24GB GPU was used for the code; Pytorch Lightning framework was used for the algorithm development. 
\footnote{Code is available at \url{https://github.com/Acrophase/Sleep_Staging_KD} }

\section{RESULTS}

We evaluated the results from the distillation model against their respective baseline model, given that the principal intention of this work is to demonstrate the potency of KD. The performance of our distillation models on the holdout test data is as shown in Table.\ref{Result_Table}. Both weighted-F1-score and accuracy metrics displayed increments in performance for the proposed distillation method FB+RB+WCE as well as ablation methods FB+WCE and RB+WCE for both 4-class sleep staging and 3-class sleep staging. The best performing model for 4-class was the RB+WCE model, where the weighted-F1 score improved from \textbf{0.451} of ECG baseline to \textbf{0.512} (weighted-F1 showed improvement by \textbf{14.30 \%}, Accuracy improved by 15.6 \%). FB+WCE was the best performing model for 3-class, with weighted-F1 improved from \textbf{0.583} of ECG baseline to \textbf{0.661} ( weighted-F1 showed improved by \textbf{13.41 \%}, Accuracy improved by 18.1 \%). However, the ECG baseline model was outperformed by all the distillation models, which corroborates the incorporation of KD in ECG based sleep staging.

Insight into the sleep stage-wise performances of distillation approaches is given in Table.\ref{Classwise_Table}. In 4-class staging, Light sleep(L) showed the best improvement, where RB+WCE ablation method improved weighted-F1 to \textbf{0.611} from \textbf{0.473}. The best improvement was observed for the NREM class in 3-class staging, with weighted-F1 improved from \textbf{0.771} to \textbf{0.652} by the FB+WCE method. Noticeably, other classes have underachieved marginally. This is potentially owing to class imbalance resulting in imprecise feature training. However, improvement across all classes in both 3-class and 4-class was observed in the proposed FB+RB+WCE distillation, thus exhibiting relatively robust feature learning against the class imbalance.

\section{DISCUSSION}

\begin{table}[t]
\caption{Performance of KD and its components
\label{Result_Table}}
\begin{minipage}{\linewidth}
\centering
\begin{tabularx}{\linewidth}{@{} C{.55in} C{1in} C{.55in} C{.55in} @{}}\toprule[1.5pt]
\hline
\textbf{Sleep Stages} & \multicolumn{1}{c}{\textbf{Model}} & \multicolumn{1}{c}{\textbf{F1-weighted}} & \textbf{Accuracy} \\ \hline
\multirow{5}{*}{W-R-L-D} & EEG Base & 0.85          & 0.85          \\
                         & ECG Base & 0.45          & 0.44          \\
                         & RB + WCE      & \textbf{0.51} & \textbf{0.51} \\
                         & FB + WCE      & 0.50          & 0.50          \\
                         & FB + RB + WCE & 0.50          & 0.49          \\ \hline
\multirow{5}{*}{W-R-N}   & EEG Base & 0.90          & 0.90          \\
                         & ECG Base & 0.58          & 0.56          \\
                         & RB + WCE      & 0.61          & 0.60          \\
                         & FB + WCE      & \textbf{0.66} & \textbf{0.66} \\
                         & FB + RB + WCE & 0.64          & 0.63          \\ \hline
\bottomrule[1.25pt]
\end{tabularx}\par
\bigskip
W:Wake, R:REM, L:Light Sleep, D:Deep Sleep, N:Non-REM Sleep
\end{minipage}
\end{table}

\begin{figure*}[t]
    \centering
      {
      \centering
      \includegraphics[width=1.0\textwidth, height=0.51\textwidth]{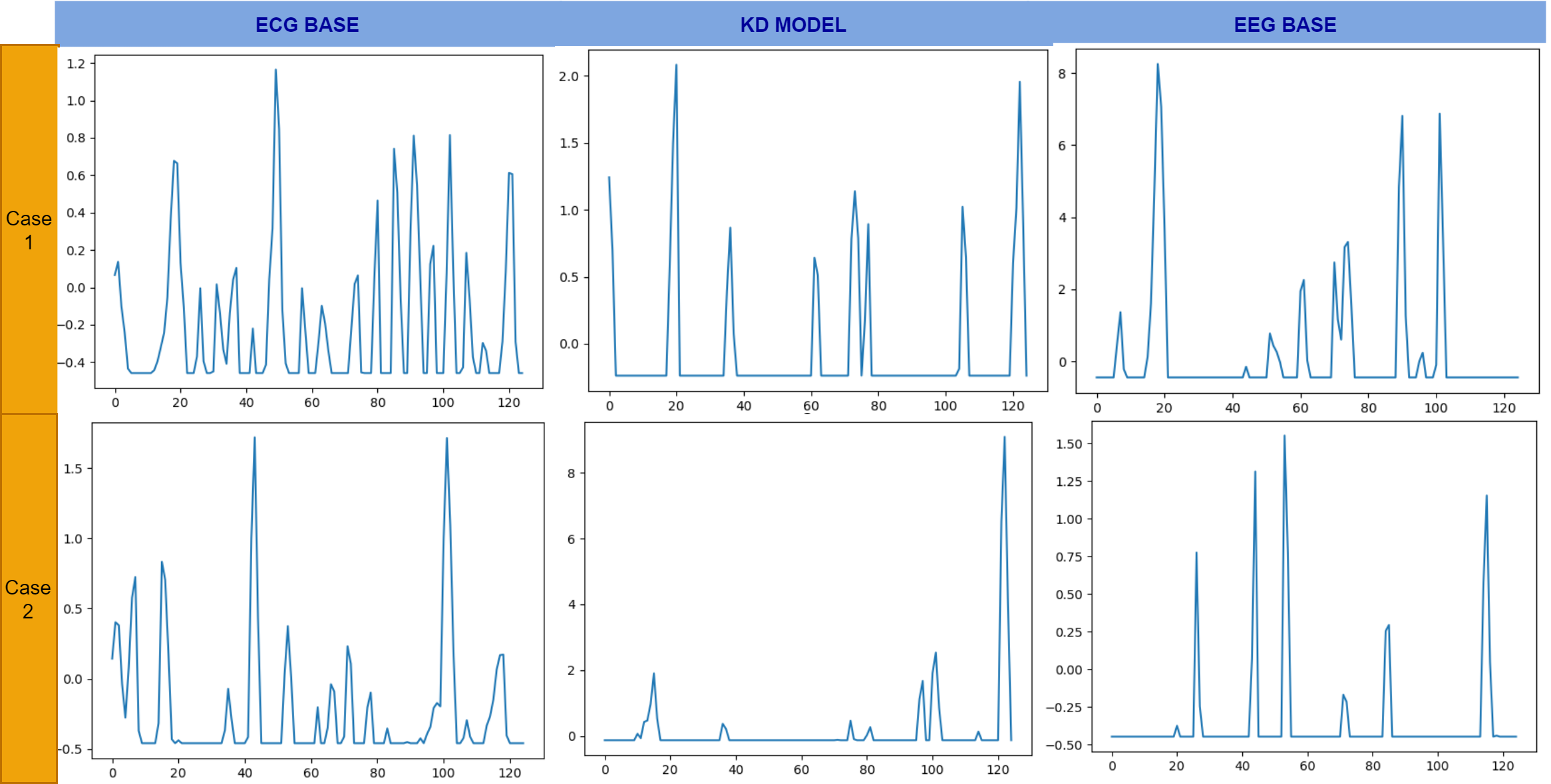}
}
    
    \caption{Comparison of bottleneck layer features for Knowledge Distillation. 
    case1: KD model correct, ECG base incorrect; case2: ECG baseline correct, KD model incorrect.}
    \label{fig:Feature_plots}
\end{figure*}

The above-indicated results demonstrate that the potency of the knowledge distillation as the proposed distillation model for ECG-based sleep staging remarkably outperformed the ECG baseline. Nevertheless, combining Feature-Based (FB) distillation and Response Based (RB) distillation need not necessarily increase the performance over independent usage of these distillation frameworks. This emphasizes the intricacy of the interaction between these two components of KD and suggests the requirement of independent optimization of the two modes of KD.

Figure.\ref{fig:Feature_plots} illustrates the feature learning in the bottleneck layer of the architecture, which represents the most compressed form of features. Although the bottleneck features themselves are not illustratable, a comparative approach assists in analyzing the functioning of KD. The figure shows two scenarios of a sleep stage prediction for 4-class sleep staging, comparing the ECG base model, the KD model and the EEG base model;
\begin{itemize}
    \item
    case 1:  ECG baseline mispredicts the sleep stage, but the proposed KD model predicts the sleep stage accurately
    \item
    \vspace{0.4em}
    case 2:  ECG baseline predicts the sleep stage accurately, but the KD model mispredicts the sleep stage. 
\end{itemize}
It is evident that distilled model's bottleneck features are comparable to that of the EEG base model's feature in case 1, which shows distinguished feature learning resulting in performance improvement. However, in case 2, misguided feature learning can be observed from the difference between the distilled model's feature and both EEG and the ECG baseline features, which were predicted accurately. This could be, to some extent, attributed to asymmetric distillation as a result of the class imbalance in the data, as observed in Table \ref{Classwise_Table}.

The methods presented here work towards producing a less-invasive alternative to sleep studies by removing cumbersome EEG electrodes, which can be prone to reduced SNR through a patient's restless sleep, and replacing them with ECG electrodes through the help of
KD. While the potential to exchange EEG for ECG signals is novel, diagnostic sleep studies still require other sensor measures per AASM guidelines, such as airflow, breathing effort, EMG from upper/lower limbs, and a pulse oximeter for oxygen saturation content. Our future efforts will be directed towards the number of sensors placed on a patient for sleep studies to improve patient comfort and their quality of sleep during the overnight study, thus optimizing for a better trade-off between comfort and accuracy.

\begin{table}[t]
\caption{KD methods class wise Results} \label{Classwise_Table}
\begin{minipage}{\linewidth}
\centering
\begin{tabularx}{\linewidth}{@{} C{1in} C{.25in} *7X @{}}\toprule[1.5pt]
\hline
\multicolumn{1}{c}{\multirow{2}{*}{}} &
  \multicolumn{4}{|c}{\textbf{4 class F1 score}} &
  \multicolumn{3}{|c}{\textbf{3 class F1 score}} \\ 
\multicolumn{1}{c}{} &
  \multicolumn{1}{|c}{\textbf{W}} &
  \multicolumn{1}{c}{\textbf{L}} &
  \multicolumn{1}{c}{\textbf{D}} &
  \textbf{R} &
  \multicolumn{1}{|c}{\textbf{W}} &
  \multicolumn{1}{c}{\textbf{N}} &
  \textbf{R} \\ \hline
\textbf{EEG Base} & 0.89 & 0.86          & 0.81 & 0.82 & 0.89 & 0.93          & 0.80 \\
\textbf{ECG Base} & 0.57 & 0.47          & 0.30 & 0.40 & 0.51 & 0.65          & 0.40 \\
\textbf{RB + WCE}      & 0.54 & \textbf{0.61} & 0.31 & 0.35 & 0.53 & 0.70          & 0.34 \\
\textbf{FB + WCE}      & 0.54 & 0.57          & 0.34 & 0.40 & 0.52 & \textbf{0.77} & 0.37 \\
\textbf{FB + RB + WCE} & 0.57 & 0.56          & 0.29 & 0.41 & 0.52 & 0.73          & 0.39 \\ \hline
\bottomrule[1.25pt]
\end{tabularx}\par
\bigskip
W:Wake, R:REM, L:Light Sleep, D:Deep Sleep, N:Non-REM Sleep
\end{minipage}
\end{table}

In spite of the promising improvement in performance brought about by KD, we identified a few limitations in this study. Firstly, using temporal models like Long Short Term Memory networks (LSTM) can improve the baseline ECG model utilized in this paper because of their ability to identify sparsely distributed features over time. Furthermore, using additional unobtrusive or less obtrusive modalities like respiratory signal and ECG have improved sleep staging performance. Previous works \cite{sridhar2020deep}\cite{li2018deep} have been trained on an extensively large dataset (>4000 records) which achieved noteworthy results on ECG based sleep staging, whereas our study used a relatively compact dataset. This shows that the choice of the dataset used in sleep staging studies cannot be undermined. Future work would involve exploring the components of KD to optimize for ECG signals as well as incorporating the benefits of KD to more optimized DL architectures, ultimately boosting the overall performance.  

\section{CONCLUSION}

This study expands the present knowledge in sleep staging from ECG by making the following contributions.
\begin{itemize}
    \item 
    Proposed usage of single modality, single-lead ECG signal, for sleep staging, minimizing the obtrusiveness and making it suitable for point of care setting.
    \item 
    Demonstration of the viability of a KD framework for two different morphological signals, resulting in improved performance of ECG-based sleep staging with knowledge assistance via EEG features for the same task.
    \item
    Analysis of the individual components of the KD by providing comparative analysis from the bottleneck layer features gave insights into the rationale for the performance improvement.
\end{itemize}

\addtolength{\textheight}{-12cm}   


\bibliographystyle{ieeetr}
\bibliography{ref}

\end{document}